\documentclass[journal]{IEEEtran}

\usepackage{cite}

\usepackage{color}

\usepackage{amsmath}

\usepackage{algorithm}
\usepackage{algorithmic}
\usepackage{bm}
\usepackage{latexsym}
\usepackage{amsthm}
\usepackage{url}
\usepackage{amsfonts}
\usepackage{amssymb}
\usepackage{indentfirst}
\usepackage{float}

\ifCLASSINFOpdf
   \usepackage[pdftex]{graphicx}
   \graphicspath{{../pdf/}{../jpeg/}}
   \DeclareGraphicsExtensions{.pdf,.jpeg,.png}
\else
   \usepackage[dvips]{graphicx}
   \graphicspath{{../eps/}}
   \DeclareGraphicsExtensions{.eps}
\fi
\usepackage{array}
\usepackage{multirow}

\ifCLASSOPTIONcompsoc
  \usepackage[caption=false,font=normalsize,labelfont=sf,textfont=sf]{subfig}
\else
  \usepackage[caption=false,font=footnotesize]{subfig}
\fi

\begin{document}
%
\title{Designing and Training of A Dual CNN for Image Denoising}
%
%
%

\author{Chunwei Tian,
        Yong Xu$^*$, \emph{Senior Member}, \emph{IEEE},
        Wangmeng Zuo, \emph{Senior Member}, \emph{IEEE},
        Bo Du, \emph{Senior Member}, \emph{IEEE},
        Chia-Wen Lin$^{*}$, \emph{Fellow}, \emph{IEEE},
        and David Zhang, \emph{Life Fellow}, \emph{IEEE}
\thanks{This work was supported in part by the National Nature Science Foundation of China Gant No. 61876051 and in part by the Shenzhen Key Laboratory of Visual Object Detection and Recognition under Grant No. ZDSYS20190902093015527. (Corresponding author: Yong Xu (Email: yongxu@ymail.com) and Chia-Wen Lin (Email: cwlin@ee.nthu.edu.tw))}
\thanks{Chunwei Tian and Yong Xu are with the Bio-Computing Research Center, Harbin Institute of Technology, Shenzhen, and Shenzhen Key Laboratory of Visual  Object Detection and Recognition, Shenzhen, 518055, Guangdong, China. Yong Xu is also with the Peng Cheng Laboratory, Shenzhen, 518055, China. (Email: chunweitian@163.com; yongxu@ymail.com)}
\thanks{Wangmeng Zuo is with the School of Computer Science and Technology, Harbin Institute of Technology, Harbin, 150001, Heilongjiang, China. He is also with the Peng Cheng Laboratory, Shenzhen, 518055, China. (Email: wmzuo@hit.edu.cn)}
\thanks{Bo Du is with the School of Computer Science, Wuhan University, Wuhan, 430072, Hubei, China. (Email: remoteking@whu.edu.cn)}
\thanks{Chia-Wen Lin with the Department of Electrical Engineering and the Institute of Communications Engineering, National Tsing Hua University,
Hsinchu, Taiwan (Email: cwlin@ee.nthu.edu.tw)}
\thanks{David Zhang is with the School of Science and Engineering, The Chinese University of Hong Kong (Shenzhen), Shenzhen, 518172, Guangdong, China. And he is also with the Shenzhen Institute of Artificial Intelligence and Robotics for Society, Shenzhen, China
(Email: davidzhang@cuhk.edu.cn)}}

\maketitle

\begin{abstract}
Deep convolutional neural networks (CNNs) for image denoising have recently attracted increasing research interest. However, plain networks cannot recover fine details for a complex task, such as real noisy images. In this paper, we propose a \textbf{Du}al \textbf{de}noising \textbf{Net}work (DudeNet) to recover a clean image. Specifically, DudeNet consists of four modules: a feature extraction block, an enhancement block, a compression block, and a reconstruction block. The feature extraction block with a sparse mechanism extracts global and local features via two sub-networks. The enhancement block gathers and fuses the global and local features to provide complementary information for the latter network. The compression block refines the extracted information and compresses the network. Finally, the reconstruction block is utilized to reconstruct a denoised image. The DudeNet has the following advantages: (1) The dual networks with a parse mechanism can extract complementary features to enhance the generalized ability of
denoiser. (2) Fusing global and local features can extract salient features to recover fine details for complex noisy images. (3) A Small-size filter is used to reduce the complexity of denoiser. Extensive experiments demonstrate the superiority of DudeNet over existing current state-of-the-art denoising methods.
\end{abstract}


\section{Introduction}
\IEEEPARstart{I}{mage} denoising is a long-standing problem in the field of low-level computer vision \cite{he2015hyperspectral,he2018hyperspectral}. In general, it can be applied to recover a high-quality image (also treated as latent clean image) $x$ via the degradation model $y = x + v$, where $y$ denotes a corrupted (noisy) image  and $v$ is additive white Gaussian noise with standard deviation of $\sigma$. In  Bayesian inference viewpoint, prior knowledge has an important effect on image denoising \cite{xu2015patch}. For instance, a weighted nuclear norm minimization (WNNM) \cite{gu2014weighted} utilizes singular values to derive different weights as solutions. Then, WNNM deals with image denoising based on nonlocal self-similarity. The block-matching and 3-D filtering (BM3D) method \cite{dabov2007image} combines 3D data and sparsity prior to tackle image denoising problems. The simultaneous use of signal processing techniques and prior is beneficial to image processing applications \cite{malfait1997wavelet}. To improve the efficiency of denoising, dictionary learning techniques were developed to suppress  noise \cite{mairal2009non}.

Although prior-based methods can achieve promising denoising performance, they are faced with the challenges of manually-set parameters and complex optimal algorithms. To address these problems, various discriminative learning methods were proposed to train the image prior models. For instance, Schmidt et al. proposed the cascade of shrinkage fields (CSF) model \cite{schmidt2014shrinkage} to recover images by using the unrolled half-quadratic optimization. Chen et al. designed the trainable nonlinear reaction diffusion (TNRD) method \cite{chen2016trainable} to train a denoising model based on an image prior of field of experts with gradient descent inference. Although these methods perform well for image denoising, their applications are limited by the used priors. Moreover, they require several hand-tuned parameters to obtain the optimal parameters \cite{zhang2017beyond}. In addition, the above methods are only useful for a certain noise level, making them ineffective for blind denoising.

The recently proposed CNNs have largely advanced image denoising \cite{tian2019deep}.  Zhang et al. presented denoising convolutional neural network (DnCNN) \cite{zhang2017beyond}, which uses residual learning (RL) and batch renormalization (BN) \cite{ioffe2015batch} to remove noise. Specifically, DnCNN first applies a single model to handle multiple applications, e.g., image denoising, super-resolution, and deblocking. The effectiveness of these multi-purpose methods relies on the
core structure used. If their core structure  cannot well recover clean image details for dynamic or complex tasks, such as
real-world corrupted images and blind noise, these methods may not perform well \cite{pan2018learning}.

To tackle these problems, we propose a \textbf{Du}al \textbf{de}noising \textbf{Net}work (DudeNet), with a lower computational cost, as shown in Fig. 1. DudeNet consists of four
components: a feature extraction block (FEB), an enhancement block (EB), a compression block (CB), and a reconstruction block (RB).  Specifically, FEB with a sparse mechanism first extracts global and locale features from a given noisy image. Then, to gradually enhance residual information, the EB fuses the global and local  features to offer complementary information to the later network through a two-phase mechanism. Next, CB is stacked to distill the residual image obtained and reduce the number of local parameters. Finally, RB reconstructs the latent clean image from residual and noisy images.

The proposed DudeNet has several merits:

(1) The proposed dual networks with a sparse mechanism can extract diverse features to boost the generalized ability of denoiser for addressing complex tasks, such as real-world corrupted images and blind noise.

(2) Fusing global with local features can obtain salient features to recover fine details, which can consolidate dual networks to tackle complex denoising tasks.

(3) A small-sized filter is used to reduce the complexity of denoiser.

The rest of this paper is organized as follows. Section II surveys several works relevant to our proposed method. Section III elaborates the proposed DudeNet. Section IV gives comprehensive experimental results with the proposed DudeNet and in-depth analyses. Conclusions are drawn in Section V.

\section{Related work}
\subsection{Deep CNNs for image denoising}
Due to the strong expressive ability and fast speed of deep CNNs, many CNN-based denoising methods have become popular for low-level vision tasks \cite{tian2020image}. Zhang et al. presented a fast and flexible denoising network as well as FFDNet based on noise level maps and noisy images \cite{zhang2018ffdnet} for blind denoising.

To better make a tradeoff between efficiency and specialized task, Zhang et al. proposed image restoration CNN (IRCNN) \cite{zhang2017learning} method by combining a discriminative learning method and model-based optimization to predict a clean image. To facilitate training, Liu et al. proposed a deep multi-level wavelet CNN (MWCNN) \cite{liu2018multi}, which fused a U-Net architecture \cite{ronneberger2015u} and wavelet to extract frequency features for image restoration tasks. Tai et al. proposed a deep-architecture persistent memory network (also named MemNet) \cite{tai2017memnet} to recover high quality images, which was composed of recursive and gate units to dig into more accurate information. Mao et al. developed a deeper 30-layer residual encoder-decoder network (RED30) \cite{mao2016image}, consisting of numerous convolutions and subsequent transposed convolutions, to obtain clearer image. Although, some methods have obtained excellent performance in image restoration, they depended on main structures. When main structures recovered well details, their performance were perfect. However, main structures cannot recovered well for varying or complex tasks, such as blind noise and real noisy image, these methods did not perform well \cite{pan2018learning}. To address this issue, Pan et al. \cite{pan2018learning} proposed dual convolution neural networks to extract complementary features for enhancing the recovered details in low-level vision task. Motivated by that, we use the dual CNNs to remove the noise, especially varying noisy images (i.e., corrupted images and blind noise from real-world applications).

\subsection{Deep CNNs based modules or blocks for image denoising}
Owing to end-to-end connection architectures, CNNs with flexible plugins are used far in many tasks, i.e., image \cite{tian2020attention}, video \cite{yuan2020learning} and text applications \cite{duan2018attention}. Specifically, modules or blocks in CNNs are used in low-level computer vision, especially, image super-resolution \cite{tian2020coarse,ahn2018fast} and denoising \cite{tian2019image}. Deep CNNs based on modules mainly divided into two categories: improving the performance, accelerating the speed.

For the first aspect, scholars mainly fused obtained multiple features to enhance the expressive abilities of CNNs. For example, deep boosting framework (DBF) \cite{chen2018deep} used feature extraction, feature integration and reconstruction blocks to suppress noise. Specifically, feature integration block used multiple concatenation operations to fuse features. A cascading residual network (CARN) \cite{ahn2018image} integrated features via repeatedly cascading residual blocks for image super-resolving. Residual dense network (RDN) \cite{zhang2018residual} repeatedly fused global and local features through recursively using residual blocks to improve image-super-resolution performance.

For the second aspect, compressing network was very common way to improve the speed of network. For example, a lightweight feature fusion network (LFFN) \cite{yang2019lightweight} exploited spindle blocks to reduce convolutional kernel size and compress trained model. Adaptive weighted super-resolution network (AWSRN) \cite{wang2019lightweight} exploited adaptive weighted multi-scale (AWMS) module to clip convolutions of small contribution. An information distillation network (IDN) \cite{hui2018fast} used three blocks (i.e., a
information extraction block, information distillation block, a construction block) to distill obtained features. Specifically, information distillation block used group convolutions and convolutional kernel of $1 \times 1$ to reduce parameters of network and computational costs.

These methods above had good effect on image super-resolution or denoising in performance or efficiency. Thus, we propose DudeNet based on blocks to narrow the differences between denoising efficiency and performance to remove the noise in this paper. The detailed information of DudeNet will be shown in Section III.

\section{Proposed Method}

\subsection{Network Architecture}
The proposed DudeNet, as illustrated in Fig. 1, is composed of  four parts: a feature extraction block (FEB), an enhancement block (EB), a compression block (CB), and a reconstruction block (RB). FEB with a sparse mechanism is designed to extract diverse features as well as reduce the depth of network. EB is used to enhance the extracted features by fusing the features two sub-networks, which is particularly useful for images corrupted by unknown types of noise, such as many real-world corrupted image and blind noise. CB compresses the network so as to reduces computational cost. Finally RB is used to reconstruct a clean image.

Specifically, FEB contains two sub-networks, namely FEBnet1 and FEBnet2, respectively, where FEBnet1 includes a sparse mechanism.  We denote the input of DudeNet as  $Y$ and its output as $X$. The two 16-layer sub-networks are exploited to extract two diverse feature maps from an input noisy image. The process of FEB can is expressed as
\begin{equation}
    FE{B_i} = {F_i}(Y),i = 1,2,
\end{equation}
where ${F_i}(Y)$ and $FEB_i$ denote the feature extraction function and the extracted features of the $i$th network, respectively.

EB contains two sections: enhancement block1 (EB1) and enhancement block2 (EB2).  The output of FEB is fed into EB1 that fuses the two diverse features from the above two sub-networks of FEB through a chain mode as follows:
\begin{equation}
    O_{E1} = {E}(FE{B_1},FE{B_2}),
\end{equation}
where $E$ represents the functions of EB1 and EB2, and $O_{E1}$ denotes the output of EB1.

The CB consists of three compression blocks (CB1), compression block2 (CB2) and compression block3 (CB3). Note that, CB1 is integrated in FEBnet1.  CB2, placed in between EB1 and EB2, is used to refine the extracted features as described below:
\begin{equation}
    O_{CB2} = {C_{1}}(O_{E1}),
\end{equation}
where $C_{1}$ denotes the functions of CB1, CB2 and CB3, respectively. $O_{CB2}$ is the output of CB2.

Following CB2, EB2 is used to obtain complementary information as formulated below:
\begin{equation}
    O_{E2} = {E}(O_{CB2},Y),
\end{equation}
where $O_{E2}$ stands for the output of EB2.

Then, CB3 further processes the output of EB2 as follows:
\begin{equation}
    O_{CB3} = {C_{1}}(O_{E2}),
\end{equation}
where $O_{CB3}$  presents the output of CB3.

Finally, RB is utilized to construct the latent clean version of input $Y$ from the obtained residual features $O_{CB3}$ using the following residual operation:
\begin{equation}
    X = Y - O_{CB3},
\end{equation}
where `$-$' denotes the residual operation.
\begin{figure*}[!htbp]
\centering
\subfloat{\includegraphics[width=7in]{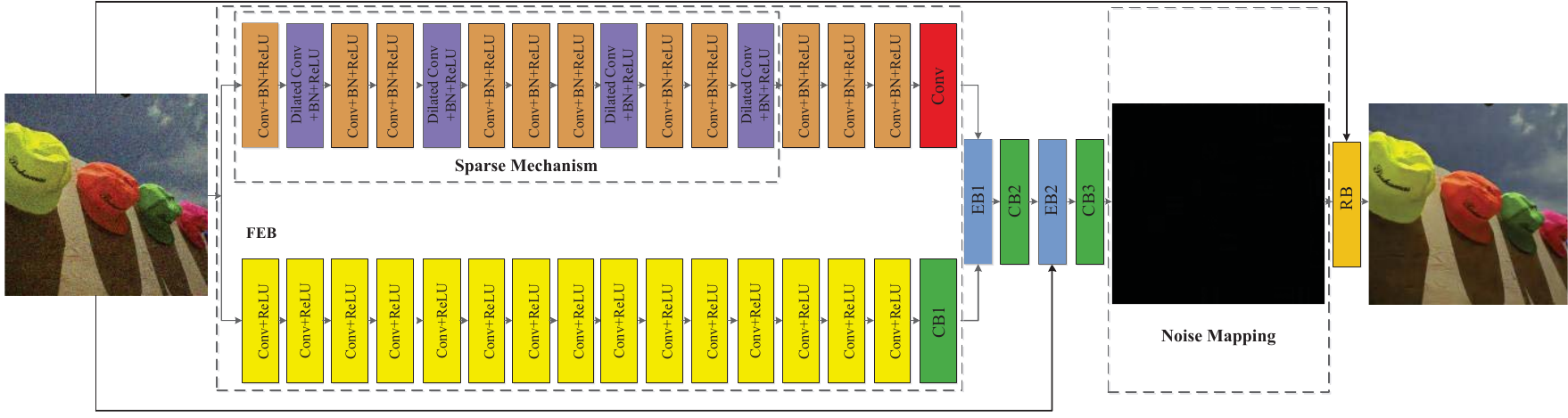}
\label{Fig:a}}
\hfil
\caption{Network architecture of the proposed DudeNet, that is composed of  four parts: a feature extraction block (FEB), an enhancement block (EB), a compression block (CB), and a reconstruction block (RB).}
\end{figure*}

\subsection{Feature Extraction Block (FEB)}
FEB is used to extract reliable visual features. The FEB consists of the first and second networks. Specifically, the first network can be split into three modules: Conv+BN+ReLU, Dilated Conv+BN+ReLU and Conv. Specifically, Dilated Conv+BN+ReLU performs dilated convolutions \cite{yu2015multi}, batch normalization \cite{ioffe2015batch} and rectified linear units \cite{krizhevsky2012imagenet}  in sequence in the proposed sparse mechanism. Conv+BN+ReLU makes  effect on the 1st, 3th, 4th, 6th, 7th, 8th, 10th, 11th, 13th, 14th and 15th layers, whereas Dilated Conv+BN+ReLU is designed in the 2th, 5th, 9th and 12th layers, and Conv forms the 16th layer. All filter sizes of the convolutional layers are set to $3 \times 3$. The sizes of 2--16 layers are $64 \times 3 \times 3 \times 64$. The sizes of the 1th layer is $c \times 3 \times 3 \times 64$, where $c$
is the channel number, where $c = 1$ and $c = 3$ mean that the input noisy images are  grayscale and color, respectively. Additionally, layers 2--12 comprise the sparse mechanism in FEBnet1. In \cite{yang2011robust}, it was shown that sparsity magnifies the effect of a small amount of big energy points. Motivated by this, we propose a sparse mechanism  in FEBnet1 of FEB. Specifically, the 2th, 5th, 9th and 12th layers can capture rich contextual information, namely big energy points, through a series of dilated convolutions with a dilation factor of 2. The other layers of FEBnet1 perform common convolutions to extract relatively fewer features than the dilated convolutional layers, namely small energy points. The joint use of big and low energy points make uses of sparsity. Thus,  layers 2--12 in FEBnet1 are called the sparse mechanism.

As a result, the function of FEBnet1 can be formulated as follows:
\begin{equation}
    FE{B_1}{\rm{ = }}{C}({CBR_3}(S({CBR_1}(Y)))),
\end{equation}
where $CBR_1$, $S$, $CBR_3$ and $C$ represent the functions of  Conv+BN+ReLU,  designed sparse mechanism,  three Conv+BN+ReLU, and one $3 \times 3$ convolution, respectively. This is converted to ${F_1}(Y) = {FEB_1}$ via (1).

The second sub-network FEBnet2 contains two different modules: Conv+ReLU and CB1. Specifically, Conv+ReLU refers to a convolution with a filter size of $3 \times 3$, followed by ReLU. CB1 is implemented via a $1 \times 1$ convolutional layer. The sizes of layers 2--15 are all $64 \times 3 \times 3 \times 64$. For the 1th and 16th layers, their sizes are respectively designed as $c \times 3 \times 3 \times 64$ and $64 \times 1 \times 1 \times 64$, where $c$ is the channel number. The procedure of FEBnet2 can be formulated as follows:
\begin{equation}
    FE{B_2} = {C_{1}}({CR_{15}}(Y)),
\end{equation}
where $CR_{15}$ expresses the function of fifteen Conv+ReLU.
\subsection{Enhancement Block (EB)}
EB uses two parts to boost the learning function of designed network, which is suitable to noise of unknown types, such as corrupted images and blind noise in real-world applications.

The proposed EB acts between FEB and CB3, and includes two parts: EB1 and EB2. Specifically, EB1 acts between FEB and CB2. EB1 has three sections: fusion part, BN and ReLU. Firstly, the fusion part integrates two different types of features from different networks (the first and second networks) via a concatenation operation \cite{takahashi2019novel}. It is known that obtained features via dilated convolutions in the first network and CB1 in the second network are different, which results in the distributions of obtained features in the EB1 have big difference. Thus, BN is used to eliminate bad effect. Finally, ReLU is employed to convert the linear features obtained into nonlinear features. This process can be formulated as follows:.

\begin{footnotesize}
\begin{equation}
    O_{E1} = R(B(CON({C}({CBR_3}(S({CBR_1}(Y)))),{CB}({CR_{15}}(Y))))),
\end{equation}
\end{footnotesize}
where $B$ is the function of BN and $R$ expresses the activation function, ReLU. EB2, acts between CB2 and CB3, concatenating the input of DudeNet and the output of CB2, and producing important information. This process can be expressed by
\begin{equation}
O_{E2} = E(O_{CB2},Y),
\end{equation}
where $O_{E2}$ acts the CB3.
\subsection{Compression Block (CB) and Reconstruction Block (RB)}
CB is used to  distill the extracted features to them more accurate, and reduces the computational cost. It consists of three parts: CB1, CB2 and CB3. Specifically, CB1 of size $64 \times 1 \times 1 \times 64$ is placed in the 16-th layer of FEBnet2 of FEB, CB2 of size $128 \times 1 \times 1 \times c$ is placed in between EB1 and EB2, and CB3 of size $2c \times 1 \times 1 \times c$, whwer $c$ is the channel size, is placed in between EB2 and RB.  Further, CB1, CB2 and CB3 are implemented by a $1 \times 1$ convolutions, which can reduce their dimension and improve the efficiency of DudeNet, since convolutions of $1 \times 1$  are known to compress data \cite{hui2018fast}. The illustrations above have presented the CB1 and CB2. Thus, the CB3 is emphatically shown as follows.
\begin{equation}
O_{CB3} = C_{1}(O_{E{2}}),
\end{equation}
where $O_{CB3}$ denotes the residual image (regarded as noise mapping) as shown in Fig. 1. The RB uses (6) to construct the predicted clean images.

\subsection{Loss Function}
We use the following mean squared error (MSE)\cite{douillard1995iterative} as the objective function (also named loss function) to measure the discrepancy  between the predicted residual image $R(Y_j)$ and the corresponding ground-truth $Y_j - {X_j}$, where ${X_j}$ denotes the $j$th clean image.
\begin{equation}
    L(\theta ) = \frac{1}{{2N}}\sum\limits_{j = 1}^N {\left\| {R({Y_j},\theta ) - ({Y_j} - {X_j})} \right\|} _2^2,
\end{equation}
where $\theta$ represents the parameters of the trained model in DudeNet. $\{ ({Y_j},{X_j})\} _{j = 1}^N$ expresses $N$ noisy-clean image pairs. The loss function is used to restore the latent clean image  via the Adam optimizer \cite{kingma2014adam}.

\section{Experiments}
\subsection{Training datasets}
Our training data is divided into two parts: synthetic and real noisy images. The synthetic noisy images of size $180 \times 180$ include gray-level and color-level images. To create this training sets, we select the same 400 images [6] for the  synthetic noisy data.  We use the following two methods \cite{simard2003best} to augment the training data for the synthetic noisy images. (1) Bicubic interpolation with the downscale factors of 0.7, 0.8, 0.9 and 1 is applied to expand the training dataset. (2) The following eight manipulations are applied to increase the diversity of the training samples: no manipulation (i.e., the original image), $90^\circ$ counterclockwise rotation, $180^\circ$ counterclockwise rotation, $270^\circ$ counterclockwise rotation, horizontal flip, $90^\circ$ counterclockwise rotation followed by horizontal flip, $180^\circ$ counterclockwise rotation followed by horizontal flip, and $270^\circ$ counterclockwise rotation followed by horizontal flip. To make the trained model more robust, each manipulation can only be applied to one image at a time, and each image is used four times in one epoch.

As for real noisy images, we use 100 JPEG-compressed mages \cite{xu2018real} of  size $512 \times 512$ as training data, which are collected using five different digital devices:
Canon 80D, Nikon D800, Canon 600D, Sony A7 II and Canon 5D Mark II with sensors of different parameters (i.e., 800, 1,600, 3,200, 6,400, 12,800 and 25,600).  Since these real noisy images are compressed, they pose greater challenge for image denoising.

\subsection{Testing datasets}
The proposed DudeNet is tested on five public benchmark datasets: BSD68 \cite{roth2009fields}, Set12 \cite{mairal2009non}, CBSD68 \cite{roth2009fields}, Kodak24 \cite{franzen1999kodak}, and CC \cite{nam2016holistic}. In these datasets, BSD68 and Set12 contain 68 and 12 gray images of different scenes, respectively. The size of each image from the two datasets is $321 \times 481$, and $256 \times 256$, respectively. CBSD68 and Kodak24 contain 68  and 24 color natural images, respectively. The images of CBSD68 and Kodak24 are in the sizes of $321 \times 481$  and $500 \times 500$, respectively. CC contains 15 corrupted $512 \times 512$ real-world as illustrated in Figure 2 in part, which are captured by 3 digital devices: Canon 5D Mark III, Nikon D600 and Nikon D800 with three ISO values (e.g. 1,600, 3,200 and 6,400).
\begin{figure}[!htbp]
\centering
\subfloat{\includegraphics[height=3.5in,width=3.5in]{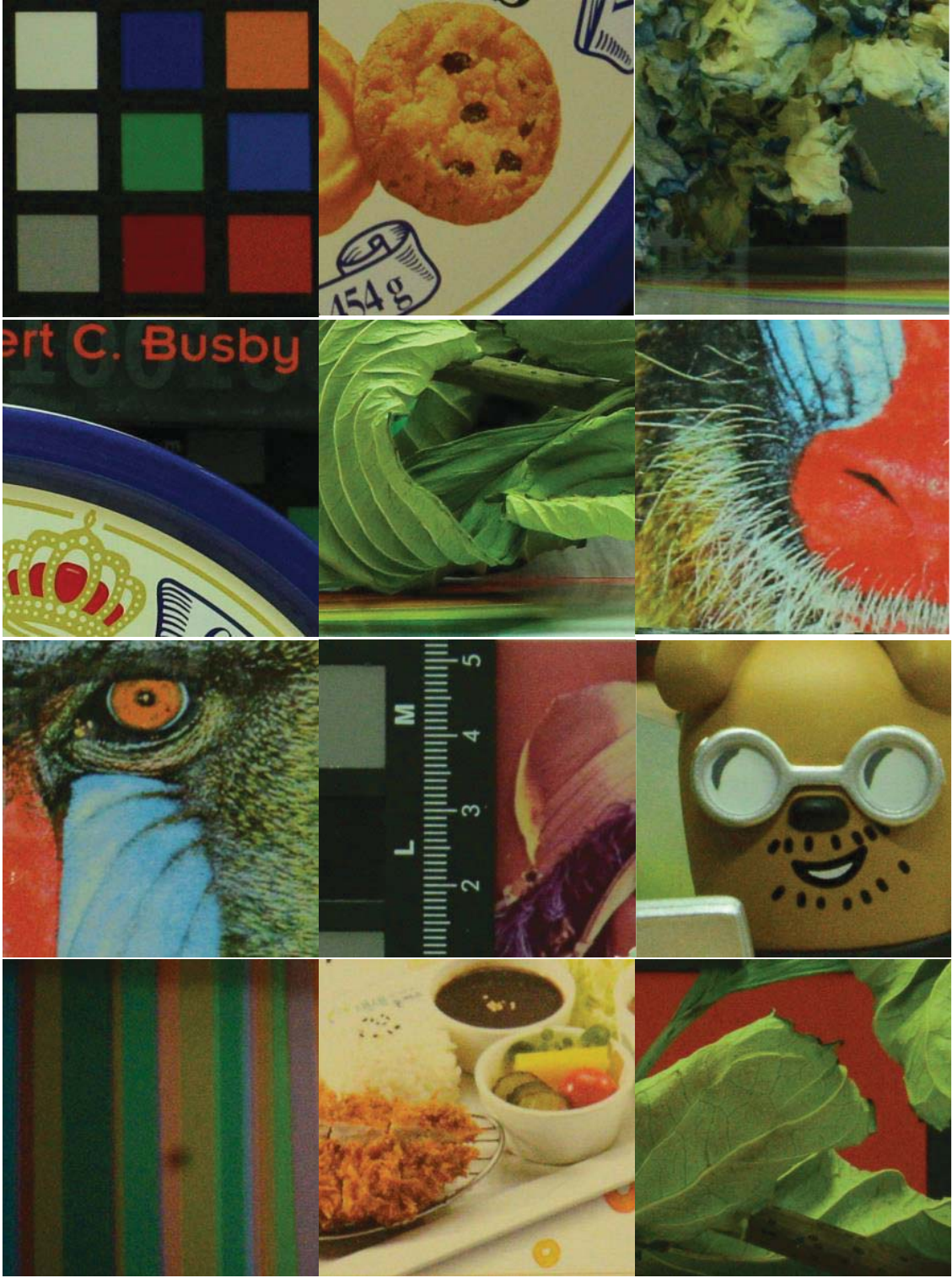}
\label{Fig:a}}
\hfil
\caption{Illustrations of 12 images in the CC dataset.}
\end{figure}
\subsection{Implementation details}
The depth of DudeNet is 18. To accelerate the speed of training, the training samples are cropped into patches of size $41 \times 41$ to train the denoising model as suggested in \cite{zoran2011learning}.

In the training phase, the initial parameter settings are as follows: a learning rate of $10^{-3}$, $\epsilon=10^{-8}$, a batch size of $128$, $\beta_1= 0.9$, $\beta_2 =0.999$  \cite{he2015delving}. The number of training epochs is 70 for the denoising models of real and synthetic noisy images, where the learning rates are set from $10^{-3}$ to $10^{-5}$.

We use Pytorch of 0.41 \cite{paszke2017pytorch} with Python of 2.7 to train and test the DudeNet model. The experiments are performed on a PC equipped with a CPU of Intel Core i7-7800X, RAM of 16G and a GPU of Nvidia GeForce GTX 1080Ti combined with an Nvidia CUDA 9.0  and a CuDNN 7.5.

\subsection{Network analysis}
\subsubsection{Design of the receptive field}
Dividing an image into patches can reduce computational cost in contrast with the entire image \cite{zoran2011learning}.  Specifically, the size of the patches is slightly larger than the receptive field size of the designed network in general. In DudeNet, the receptive field sizes of the first and second networks are $41 \times 41$ and $32 \times 32$, respectively, following \cite{yu2015multi}. It is noted that the patch size should be  greater than the receptive field size of DudeNet; otherwise, the patches cannot fit the size for the padding of network, which will degrade the performance of denoising. Considering the trade-off between computational cost and denoising performance, we take the average receptive field size of the two designed sub-networks as the receptive field size of FEB:  $(41 + 32)/2 = 36.5 \approx 37$. Thus, the overall receptive field size of DudeNet is $39 \times 39$ and the patch size is chosen as $41 \times 41$.

\subsubsection{Design, analysis and effectiveness of FEB}
To extract accurate features, we use a features fusion method to enhance the representation power of DudeNet. That is, FEB includes the first and second networks. For the first network, the proposed sparse mechanism is an important component and has the following merits. First, it can better recover the details of the latent image. Second, it can result in shallow architecture, which facilitates addressing the long-term dependency problem. Third, the shallow DudeNet has lower computational complexity. Its detailed implementations are shown in Section III. C. Specifically, this mechanism uses dilated convolutions with a large dilated factor to imitate big energy points, whereas the  convolutions with a small dilated factor are used to represent small energy points to achieve sparsity. However, the choice of big energy points is crucial. We explain the reasons from the aspects of sparsity characteristics and network design, respectively.

For sparsity, it is known that large energy points are and irregularly distributed \cite{tian2018fft}. Motivated by that, we propose the detailed requirements to find big energy points.

1) Big energy points are not successive and equidistant in CNN, which is useful to improve denoising performance. If big energy points are successive in a CNN, which would result in information loss. Specifically, memory ability of CNN is finite \cite{tao2018deep}. The latter layer in CNN need learn new content for a bigger area when it does not fully understand information of former layer. The fact is proved that `The first network with successive big energy points, CB2, CB3' has poorer performance than that of `The combination of the first network with sparse mechanism, CB2 and CB3 (FS)' as shown in Table I.  Specifically, `The first network with successive big energy points, CB2, CB3' is that the combination of CB2, CB3 and the first network with big energy points (dilated convolutions) of layers 2-5. In addition, it is noted if difference of a network in local areas is bigger, its performance is better \cite{zhang2018residual}. Thus, we do not choose equidistant big points in CNN. That is verified via `DudeNet with dilated factor of 2 in layers 2, 5, 8 and 11 in FEBnet1' and `DudeNet' in Table I.
\begin{table}[t!]
\caption{Gaussian denoising results of twelve specific models for gray noisy images. These models are trained with $\sigma  = 25$, and evaluated on the BSD68 dataset.}
\label{tab:1}
\centering
\scalebox{0.60}[0.60]{
\begin{tabular}{|c|c|}
\hline
Methods &PSNR\\
\hline
DudeNet  &29.285\\
\hline
DnCNN &29.210\\
\hline
Two DnCNNs &29.213\\
\hline
DudeNet without CB2 and EB2 &29.258\\
\hline
DudeNet without sparse mechanism, CB2 and EB2 &29.226\\
\hline
DudeNet with CB\_2 and without EB2 &29.260\\
\hline
The combination of RB, EB, CB and FEB without sparse mechanism  &29.258  \\
\hline
The combination of RB, EB1, CB and FEB without sparse mechanism  &29.257 \\
\hline
The combination of RB, EB1, CB and FEB without sparse mechanism and BN &29.116\\
\hline
The combination of the first network with sparse mechanism, CB2 and CB3 (FS) &29.067\\
\hline
The combination of FEBnet1 without sparse mechanism, CB2 and CB3  (FWS)  &28.971\\
\hline
FEBnet1 with successive big energy points, CB2, CB3 &28.991\\
\hline
FEBnet2 with CB2, and CB3  &28.985\\
\hline
The combination of each layer with a big energy points in the first network, CB2 and CB3 (ELEP) &28.474\\
\hline
DudeNet with two sparse mechanisms &29.270\\
\hline
DudeNet with dilated factor of 2 in layers 2, 5, 8 and 11 in FEBnet1 &29.259\\
\hline
DudeNet with kernel of size $3\times3$ in each layer &29.284\\
\hline
\end{tabular}}
\label{tab:booktabs}
\end{table}

2) Big energy points are not multiple, which can ensure denoising efficiency. To give an example, ELEP consumes more run-time than FS in dealing with a noisy image as shown in Table II.
\begin{table}[t!]
\caption{Comparison of Runn-time complexity of the five denoising networks for noisy images of sizes 256 $\times$ 256, 512 $\times$ 512 and 1024 $\times$ 1024.}
\label{tab:1}
\centering
\scalebox{0.70}[0.70]{
\begin{tabular}{|c|c|c|c|c|}
\hline
Methods &Device &256 $\times$ 256 &512 $\times$ 512 &1024 $\times$ 1024\\
\hline
FWS  &GPU &0.007	&0.159	&0.588 \\
\hline
FS  &GPU &0.010	&0.227	&0.873 \\
\hline
ELEP  &GPU &0.019 &0.447 &1.272 \\
\hline
DudeNet with kernel of size $3\times3$ in each layer &GPU &0.019	&0.434	&1.252 \\
\hline
DnCNN  &GPU &0.009 &0.195 &0.731\\
\hline
Two DnCNNs &GPU &0.017 &0.385 &1.217\\
\hline
DudeNet &GPU &0.018	&0.422	&1.246\\
\hline
\end{tabular}}
\label{tab:booktabs}
\end{table}
\begin{table}[t!]
\caption{Complexity analysis of different networks.}
\label{tab:1}
\centering
\scalebox{0.95}[0.95]{
\begin{tabular}{|c|c|c|c|c|}
\hline
Methods &Parameters &Gflops \\
\hline
DnCNN &0.56M &0.94 \\
\hline
Two DnCNNs &1.11M &1.87 \\
\hline
RED30 &4.13M  &10.33\\
\hline
DudeNet with kernel of size $3\times3$ in each layer &1.11M &1.87 \\
\hline
DudeNet &1.03M &1.73\\
\hline
\end{tabular}}
\label{tab:booktabs}
\end{table}

For network architecture, big energy points are successive, equidistant and multiple, which can result in low efficiency and poor denoising performance. To give an example, we assume that each layer uses dilated convolutions in the first network. That requires a padding operation in each layer in the first network, which has a lower efficiency, such as `ELEP' and `FS', as illustrated in Table II. In addition, when the size of receptive field  is greater than the patch size of the input image, the feature mapping requires zero padding, which can degrade denoising performance. This can be seen when comparing the performance of ELEP and FS in Table I. Thus, this idea is abandoned in image denoising. Further, sparse mechanism is competitive in efficiency and performance. Specifically, the FS has receptive field size of $43 \times 43$, which obtains a denoising effect in comparison with 21 layers under the kernel size of $3\times3$. However, it only has 18 layers, which reduces the depth of network to improve the denoising efficiency.

For denoising performance, we demonstrate the effectiveness of sparse mechanism by comparing `DudeNet' and `The combination of RB, EB, CB and FEB without sparse mechanism' in Table I. And `DudeNet without CB2 and EB2' outperforms `DudeNet without sparse mechanism, CB2 and EB2' as illustrated in Table I. In summary, the proposed sparse mechanism in FEBnet1 of FEB is effective. Moreover, FEBnet2 consolidates FEBnet1 to improve the denoising performance, which is explained in details in EB.

\subsubsection{Design, analysis and effectiveness of EB}
It is noticed that different networks can provide complementary information from multiple views \cite{tao2018deep}. And differences of local architectures of a CNN are bigger; its effect is better \cite{zhang2018residual}. Motivated by that, we propose an EB to enhance learning ability. Specifically, EB includes two parts: EB1 and EB2. EB1 gathers two sub-networks to improve the expressive power of the network when depth is same. It is seen that `The combination of RB, EB1, CB and FEB without sparse mechanism and BN' achieves higher peak signal to noise ratio (PSNR) than these of `FWS' and `The second network with CB2, and CB3'. `DudeNet with CB2 and without EB2' is superior to both `FS' and `The second network with CB2, and CB3'.
These prove that the combination of FEB and EB1 is very effective for image denoising. Specifically, two sub-networks use only one sparse mechanism to enlarge the difference for FEB, which is effective in Table I, such as `DudeNet with two sparse mechanisms' and `DudeNet'. That proves that dual networks with sparse mechanism can extract different features to boost the generalized ability of denoiser.

EB2 fuses local features obtained and original image information, which can also supplement the information for the first phase and later network. This is tested through comparing both `The combination of RB, EB1, CB and FEB without sparse mechanism' and `The combination of RB, EB, CB and FEB without sparse mechanism', both `DudeNet' and `DudeNet with CB\_2 and without EB2'. Meanwhile, the incorporation of BN into the FEB and EB has a positive effect on DudeNet, which is reflected in Table I. Its reasons have the following two points. Firstly, dilated convolutions in the first network can result in the distributions of obtained features are different. Secondly, obtained features from two different networks are different, which result in distribution of fused features is not same. These have na\"ive effect on image denoising. Thus, we choose BN into the first network and EB1 in FEB to address these problems, respectively. In addition, it is noted that EB1 and EB2 gathers local features and global features to boost the expressive ability, which is very suitable to unknown corrupted images, such as real-world noisy images and blind denoising.
\subsubsection{Design, analysis and effectiveness of CB}
To improve the efficiency, CB is used to discount redundancy feature information via a $1\times1$ convolution. Because CB1 is embedded into the second network of the FEB and CB3 can convert noise features into noise mapping (also called noisy image). Thus, we only prove the effectiveness of the CB2 for image denoising. That is `DudeNet with CB\_2 and without EB2' obtained higher PSRN than that of `DudeNet without CB2 and EB2' as shown in Table I. Further, it is seen that DudeNet with CB (also referred to as DudeNet) is very competitive with `DudeNet with kernel of size $3\times3$ in each layer' in performance, running time and computational cost as shown in Tables I-III. Specifically, to make the denoising result more convincible, the depth of DudeNet refers to DnCNN. However, due to the existence of EB2 and CB3, the depth of DudeNet has one layer over DnCNN. Additionally, our network is wider than DnCNN. Take into consideration these factors, we choose 'Two DnCNNs' of 18 layers as a compared method to test the denoising performance from PSNR, run-time and complexity, where `Two DnCNNs' concatenates two same DnCNNs.

For denoising result, `DudeNet' is superior to `DnCNN' and `Two DnCNNs' as shown in TABLE I. For run-time, we can see that the `DudeNet' is close to `Two DnCNNs' for a noisy image (i.e., 256 $\times$ 256, 512 $\times$ 512 and 1024 $\times$ 1024) as presented in TABLE II. For complexity, it is known that the `DudeNet' has less the number of parameters and Gflops than that of `Two DnCNNs' as described in TABLE III. Also, `DudeNet' outperforms `RED30' and `DudeNet with kernel of size 3 $\times$ 3 in each layer' in terms of the complexity of the denoising network. That shows that the shallow architecture and small filter size of the part has less computational cost and memory. As a result, our DudeNet is effective and efficient for image denoising.
\begin{table*}[!htb] \small
\caption{Average PSNR (dB) for different methods on BSD68, with various noise levels, i.e., 15, 25 and 50.}
\label{tab:1}
\centering
\scalebox{0.70}[0.70]{
\begin{tabular}{|c|c|c|c|c|c|c|c|c|c|c|c|c|c|c|c|c|c|c|}
\hline
Methods & BM3D & WNNM & EPLL & MLP & CSF & TNRD & DnCNN & IRCNN &FFDNet &ECNDNet &DnCNN-B &RED30 &MemNet &SANet &PSN-U &DudeNet & DudeNet-B  \\
\hline
$\sigma$ = 15	&31.07 &31.37 &31.21 &-  &31.24 &31.42 &\textcolor{blue}{31.73} &31.63  &31.62 & 31.71 &31.61  &- &- &31.68 &31.60 &\textcolor{red}{31.78} &31.64\\
\hline
$\sigma$ = 25	&28.57 &28.83 &28.68 &28.96 &28.74 &28.92 & \textcolor{blue}{29.23} &29.15  &29.19  &29.22 &29.16 &- &- &29.13 &29.17 &\textcolor{red}{29.29} &29.19\\
\hline
$\sigma$ = 50	&25.62 &25.87 &25.67 &26.03 &-  &25.97 &26.23 &26.19  &26.30  &26.23 &26.23 &\textcolor{red}{26.35} &\textcolor{red}{26.35} &26.10 &26.30 &\textcolor{blue}{26.31} &26.25\\
\hline
\end{tabular}}
\end{table*}
\begin{table*}[t!]
\label{tab:1}
\caption{Average PSNR (dB) for different methods on Set12, with various noise levels, i.e., 15, 25 and 50.}
\centering
\scalebox{0.70}[0.70]{
\begin{tabular}{|c|c|c|c|c|c|c|c|c|c|c|c|c|c|}
\hline
Images & C.man & House & Peppers & Starfish & Monarch & Airplane & Parrot & Lena & Barbara & Boat & Man & Couple & Average \\
\hline
\hline
Noise Level  & \multicolumn{13}{c|}{$\sigma$ = 15} \\
\hline
BM3D \cite{dabov2007image}  &31.91 &34.93 &32.69 &31.14 &31.85 &31.07 &31.37 &34.26 &\textcolor{blue} {33.10} &32.13 &31.92 &32.10 &32.37\\
\hline
WNNM \cite{gu2014weighted} &32.17 & \textcolor{red} {35.13} &32.99 &31.82 &32.71 &31.39 &31.62 &34.27 & \textcolor{red} {33.60} &32.27 &32.11 &32.17 &32.70\\
\hline
EPLL \cite{zoran2011learning} &31.85 &34.17 &32.64 &31.13 &32.10 &31.19 &31.42 &33.92 &31.38 &31.93 &32.00 &31.93 &32.14\\
\hline
CSF \cite{schmidt2014shrinkage} &31.95 &34.39 &32.85 &31.55 &32.33 &31.33 &31.37 &34.06 &31.92 &32.01 &32.08 &31.98 &32.32\\
\hline
TNRD \cite{chen2016trainable} &32.19 &34.53 &33.04 &31.75 &32.56 &31.46 &31.63 &34.24 &32.13 &32.14 &32.23 &32.11 &32.50\\
\hline
DnCNN \cite{zhang2017beyond} &\textcolor{blue} {32.61} &34.97 &33.30 & \textcolor{blue}{32.20}  &33.09 & 31.70 &31.83 & \textcolor{blue}{34.62} &32.64 & \textcolor{blue}{32.42} & \textcolor{red}{32.46} & \textcolor{blue}{32.47} & \textcolor{blue}{32.86}\\
\hline
IRCNN \cite{zhang2017learning} &32.55 &34.89 &\textcolor{blue}{33.31} &32.02 &32.82 & 31.70 & \textcolor{blue}{31.84}
&34.53 &32.43 &32.34 &32.40 &32.40 &32.77\\
\hline
FFDNet \cite{zhang2018ffdnet} &32.43 &\textcolor{blue} {35.07} &33.25 &31.99 &32.66 &31.57 &31.81 & \textcolor{blue}{34.62} &32.54 &32.38 &\textcolor{blue}{32.41} &32.46 &32.77\\
\hline
ECNDNet \cite{tian2019enhanced}  &32.56 &34.97 &33.25 &32.17	&33.11	&31.70	&31.82	&32.52	&32.41	&32.37	&32.39	&32.39	&32.81\\
\hline
DnCNN-B \cite{zhang2017beyond} &32.10 &34.93 &33.15 &32.02 &32.94 &31.56 &31.63 &34.56 &32.09 &32.35 & \textcolor{blue}{32.41} &32.41 &32.68\\
\hline
SANet \cite{zhang2019separation}  &32.38 &35.03 &33.18	&32.14	&\textcolor{blue} {33.20}	&\textcolor{blue} {31.71}	&31.89	&34.54	&32.61	&32.36	&32.38	&32.41	&32.82\\
\hline
PSN-U \cite{aljadaany2019proximal} &32.04 &35.03 &33.21 &31.94 &32.93 &31.61 &31.62 &34.56 &32.49 &32.41 &32.37 &32.43 &32.72\\
\hline
DudeNet & \textcolor{red}{32.71} & \textcolor{red}{35.13}  & \textcolor{red}{33.38}  & \textcolor{red}{32.29}
& \textcolor{red}{33.28} & \textcolor{red}{31.78} & \textcolor{red}{31.93}   & \textcolor{red}{34.66} & 32.73 & \textcolor{red}{32.46} & \textcolor{red}{32.46} & \textcolor{red}{32.49} & \textcolor{red}{32.94} \\
\hline
DudeNet-B &32.28 &35.03 &33.25 &32.12 &33.06 &31.66 &31.77 &34.58 &32.36 &32.39 &32.38 &32.41 &32.77 \\
\hline
Noise Level  & \multicolumn{13}{c|}{$\sigma$ = 25} \\
\hline
BM3D \cite{dabov2007image} &29.45 &32.85 &30.16 &28.56 &29.25 &28.42 &28.93 &32.07 &\textcolor{blue}{30.71} &29.90 &29.61 &29.71 &29.97\\
\hline
WNNM \cite{gu2014weighted} &29.64 & 33.22 &30.42 &29.03 &29.84 &28.69 &29.15 &32.24 & \textcolor{red}{31.24} &30.03 &29.76 &29.82 &30.26\\
\hline
EPLL \cite{zoran2011learning} &29.26 &32.17 &30.17 &28.51 &29.39 &28.61 &28.95 &31.73 &28.61 &29.74 &29.66 &29.53 &29.69\\
\hline
MLP \cite{burger2012image} &29.61 &32.56 &30.30 &28.82 &29.61 &28.82 &29.25 &32.25 &29.54 &29.97 &29.88 &29.73 &30.03\\
\hline
CSF \cite{schmidt2014shrinkage} &29.48 &32.39 &30.32 &28.80 &29.62 &28.72 &28.90 &31.79 &29.03 &29.76 &29.71 &29.53 &29.84\\
\hline
TNRD \cite{chen2016trainable} &29.72 &32.53 &30.57 &29.02 &29.85 &28.88 &29.18 &32.00 &29.41 &29.91 &29.87 &29.71 &30.06\\
\hline
DnCNN \cite{zhang2017beyond} &\textcolor{blue}{30.18} &33.06 &30.87   &29.41 &30.28 & \textcolor{blue}{29.13} &29.43 &32.44 &30.00 &30.21 &\textcolor{blue}{30.10} &30.12 &30.43\\
\hline
IRCNN \cite{zhang2017learning} &30.08 &33.06 &30.88 &29.27 &30.09 &29.12 &\textcolor{blue}{29.47} &32.43 &29.92 &30.17 &30.04 &30.08 &30.38\\
\hline
FFDNet\cite{zhang2018ffdnet}	&30.10	&\textcolor{red}{33.28}	&\textcolor{blue}{30.93}	&29.32	&30.08	&29.04	&29.44	&\textcolor{red}{32.57}	&30.01	&\textcolor{red}{30.25}	&\textcolor{red}{30.11}	&\textcolor{red}{30.20}	&\textcolor{blue}{30.44}\\
\hline
ECNDNet \cite{tian2019enhanced} &30.11	&33.08	&30.85	&\textcolor{blue} {29.43}	&30.30	&29.07	&29.38	&32.38	&29.84	&30.14	&30.03	&30.03	&30.39\\
\hline
DnCNN-B \cite{zhang2017beyond} &29.94 &33.05 &30.84 &29.34 &30.25 &29.09 &29.35 &32.42 &29.69 &30.20 &30.09 &30.10 & 30.36\\
\hline
SANet \cite{zhang2019separation} &30.04	&33.05	&30.83	&29.31	&30.27	&29.08	&29.34	&32.35	&30.00	&30.12	&30.00	&30.05	&30.37\\
\hline
PSN-U \cite{aljadaany2019proximal} &29.79 &33.23 &30.90 &29.30 &30.17 &29.06 &29.25 &32.45 &29.94 &\textcolor{red} {30.25} &30.05 &30.12 &30.37\\
\hline
DudeNet	&\textcolor{red}{30.23}	&\textcolor{blue}{33.24}	&\textcolor{red}{30.98}	&\textcolor{red}{29.53}	&\textcolor{red}{30.44}	&\textcolor{red}{29.14}	&\textcolor{red}{29.48}	&\textcolor{blue}{32.52}	&30.15	&\textcolor{blue}{30.24}	&30.08	&\textcolor{blue}{30.15}	&\textcolor{red}{30.52}\\
\hline
DudeNet-B &30.01 &33.15 &30.87 &29.39 &\textcolor{blue}{30.31} &29.07 &29.40 &32.42 &29.76 &30.18 &30.03 &30.06 &30.39\\
\hline
Noise Level  & \multicolumn{13}{c|}{$\sigma$ = 50} \\
\hline
BM3D \cite{dabov2007image} &26.13 &29.69 &26.68 &25.04 &25.82 &25.10 &25.90 &29.05 &\textcolor{blue}{27.22}  &26.78 &26.81 &26.46 &26.72\\
\hline
WNNM \cite{gu2014weighted} &26.45 &\textcolor{red}{30.33} &26.95 &25.44 &26.32 &25.42 &26.14 &29.25 &\textcolor{red}{27.79} &26.97 &26.94 &26.64 &27.05\\
\hline
EPLL \cite{zoran2011learning} &26.10 &29.12 &26.80 &25.12 &25.94 &25.31 &25.95 &28.68 &24.83 &26.74 &26.79 &26.30 &26.47\\
\hline
MLP \cite{burger2012image} &26.37 &29.64 &26.68 &25.43 &26.26 &25.56 &26.12 &29.32 &25.24 &27.03 &27.06 &26.67 &26.78\\
\hline
TNRD \cite{chen2016trainable} &26.62 &29.48 &27.10 &25.42 &26.31 &25.59 &26.16 &28.93 &25.70 &26.94 &26.98 &26.50 &26.81\\
\hline
DnCNN \cite{zhang2017beyond} &27.03 &30.00 &27.32 &25.70 &26.78 &25.87 &26.48 &29.39 &26.22 &27.20 & \textcolor{red} {27.24} &26.90 &27.18\\
\hline
IRCNN \cite{zhang2017learning} &26.88 &29.96 &27.33 &25.57 &26.61 &\textcolor{red}{25.89} &\textcolor{blue}{26.55} &29.40 &26.24 &27.17 &27.17 &26.88 &27.14\\
\hline
ECNDNet \cite{tian2019enhanced} &27.07	&30.12	&27.30	&\textcolor{blue}{25.72}	&26.82	&25.79	&26.32	&29.29	&26.26	&27.16	&27.11	&26.84	&27.15\\
\hline
DnCNN-B \cite{zhang2017beyond} &27.03 &30.02 &27.39 &\textcolor{blue}{25.72} &\textcolor{blue}{26.83} &\textcolor{red}{25.89} &26.48 &29.38 &26.38 &27.23 &\textcolor{blue}{27.23} &26.91 &27.21\\
\hline
SANet \cite{zhang2019separation} &26.92 &29.93	&27.27	&25.52	&26.64	&25.71	&26.18	&29.22	&26.37	&27.20	&27.11	&26.80	&27.09\\
\hline
PSN-U \cite{aljadaany2019proximal} &\textcolor{blue}{27.21} &30.21 &\textcolor{red}{27.53} &25.63 &\textcolor{red}{26.93} &\textcolor{red}{25.89} &\textcolor{red}{26.62} &\textcolor{red}{29.54} &26.56 &\textcolor{red}{27.27} &\textcolor{blue}{27.23} &\textcolor{red}{27.04} &\textcolor{red}{27.30}\\
\hline
DudeNet &\textcolor{red}{27.22}  &\textcolor{blue}{30.27} &\textcolor{blue}{27.51} &\textcolor{red}{25.88} &\textcolor{red}{26.93} &\textcolor{blue}{25.88} &26.50 &\textcolor{blue}{29.45}  &26.49 &\textcolor{blue}{27.26} &27.19 &\textcolor{blue}{26.97} &\textcolor{red}{27.30}\\
\hline
DudeNet-B &27.19 &30.11 &27.50 &25.69 &26.82 &25.85 &26.46 &29.35 &26.38 &27.20 &27.13 &26.90 &\textcolor{blue}{27.22}\\
\hline
\end{tabular}}
\end{table*}
\begin{figure}[!htbp]
\centering
\subfloat{\includegraphics[width=3.5in]{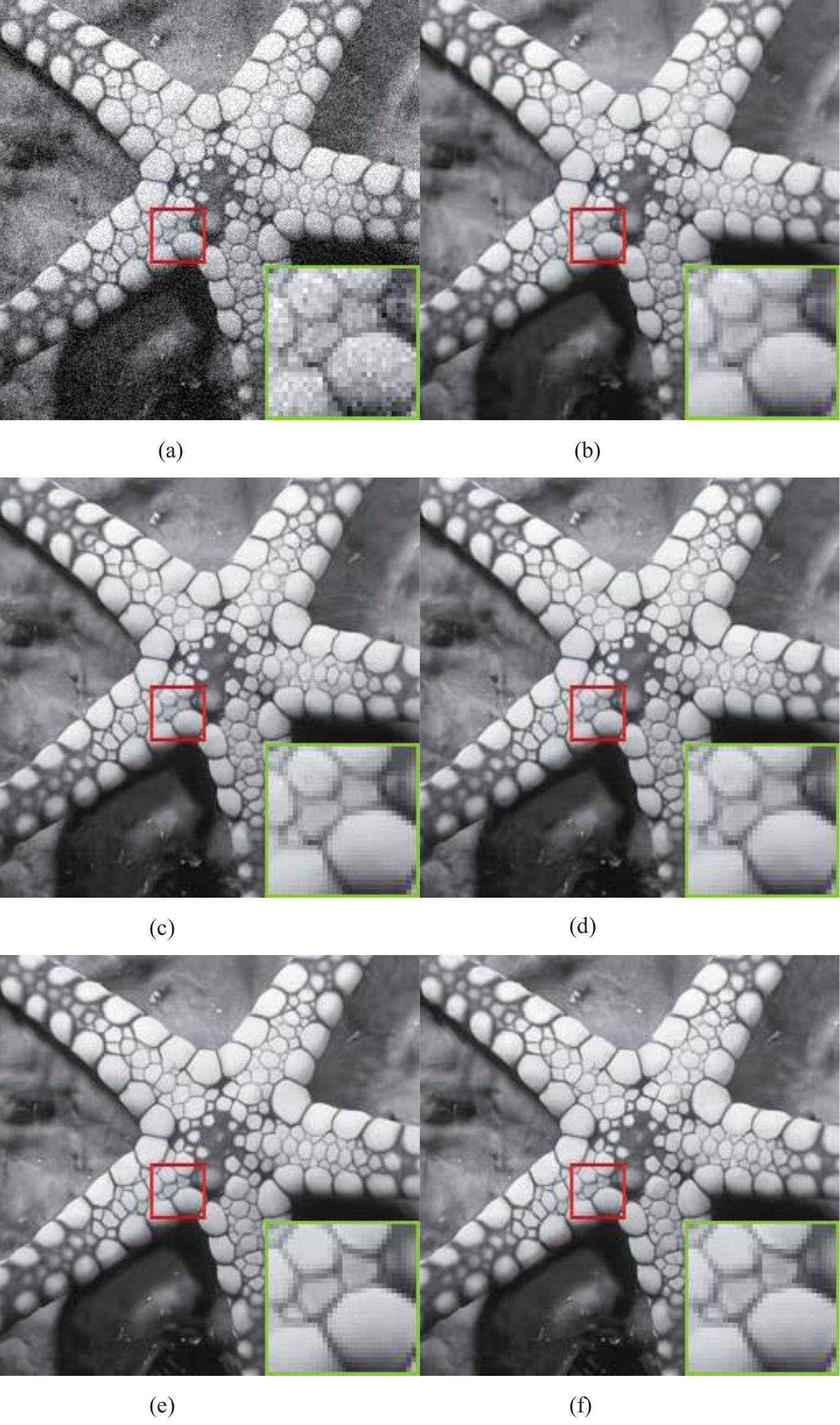}
\label{Fig:a}}
\hfil
\caption{Visual comparison for a gray Gaussian noisy image from Set12 obtained by different methods, when noise level is 15. (a) Noisy image/24.66dB, (b) BM3D/31.14dB, (c) IRCNN/32.02dB, (d) FFDNet/32.02dB, (e) DudeNet/32.29dB and (f) DudeNet-B/32.12dB.}
\end{figure}
\begin{figure}[!htbp]
\centering
\subfloat{\includegraphics[width=3.6in]{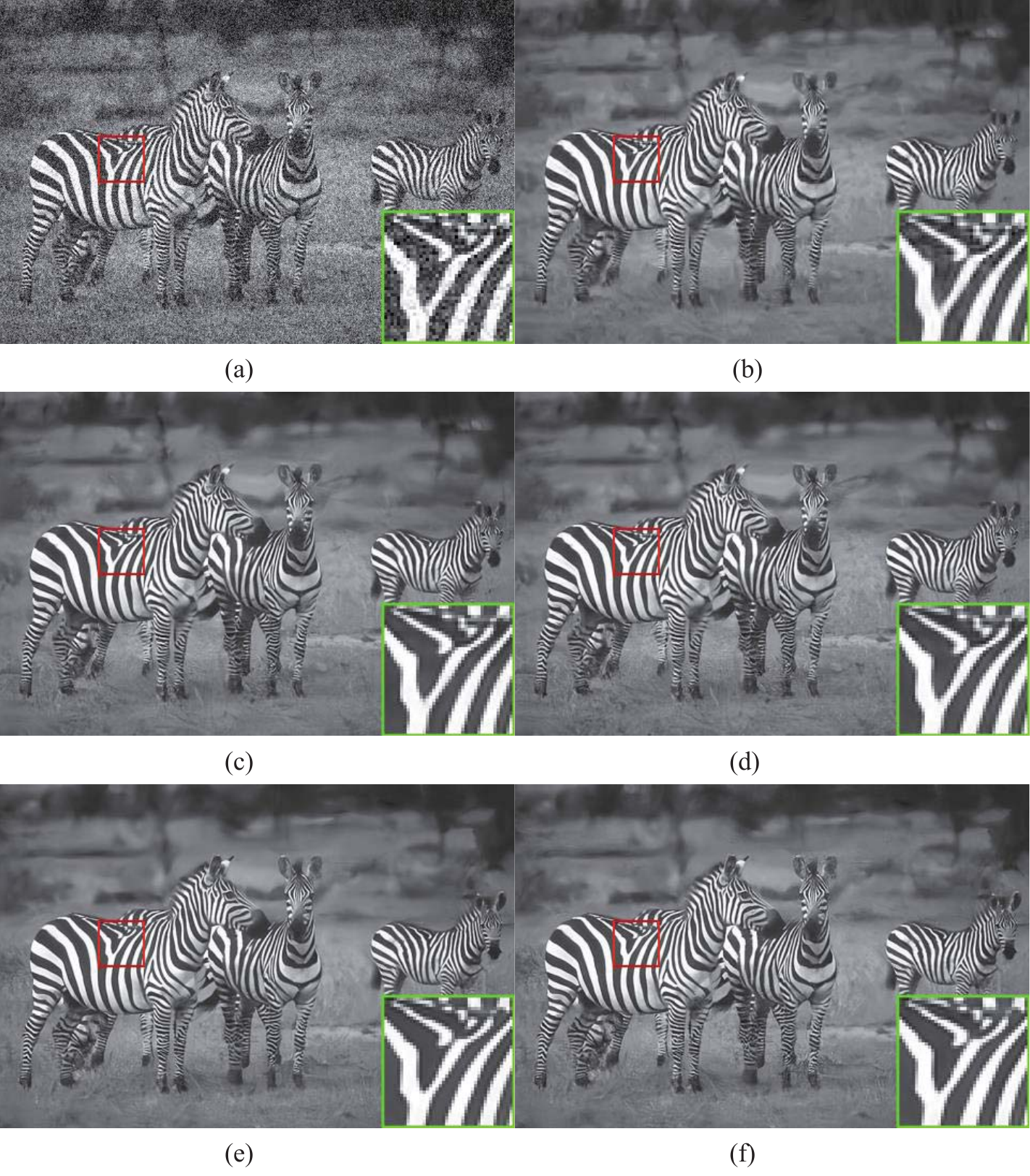}
\label{Fig:a}}
\hfil
\caption{Visual comparison for a gray Gaussian noisy image from BSD68 obtained by different methods, when noise level is 25. (a) Noisy image/20.26dB, (b) BM3D/27.56dB, (c) IRCNN/28.36dB, (d) FFDNet/28.35dB, (e) DudeNet/28.67dB and (f) DudeNet-B/28/54dB.}
\end{figure}
\begin{figure}[!htbp]
\centering
\subfloat{\includegraphics[width=3.5in]{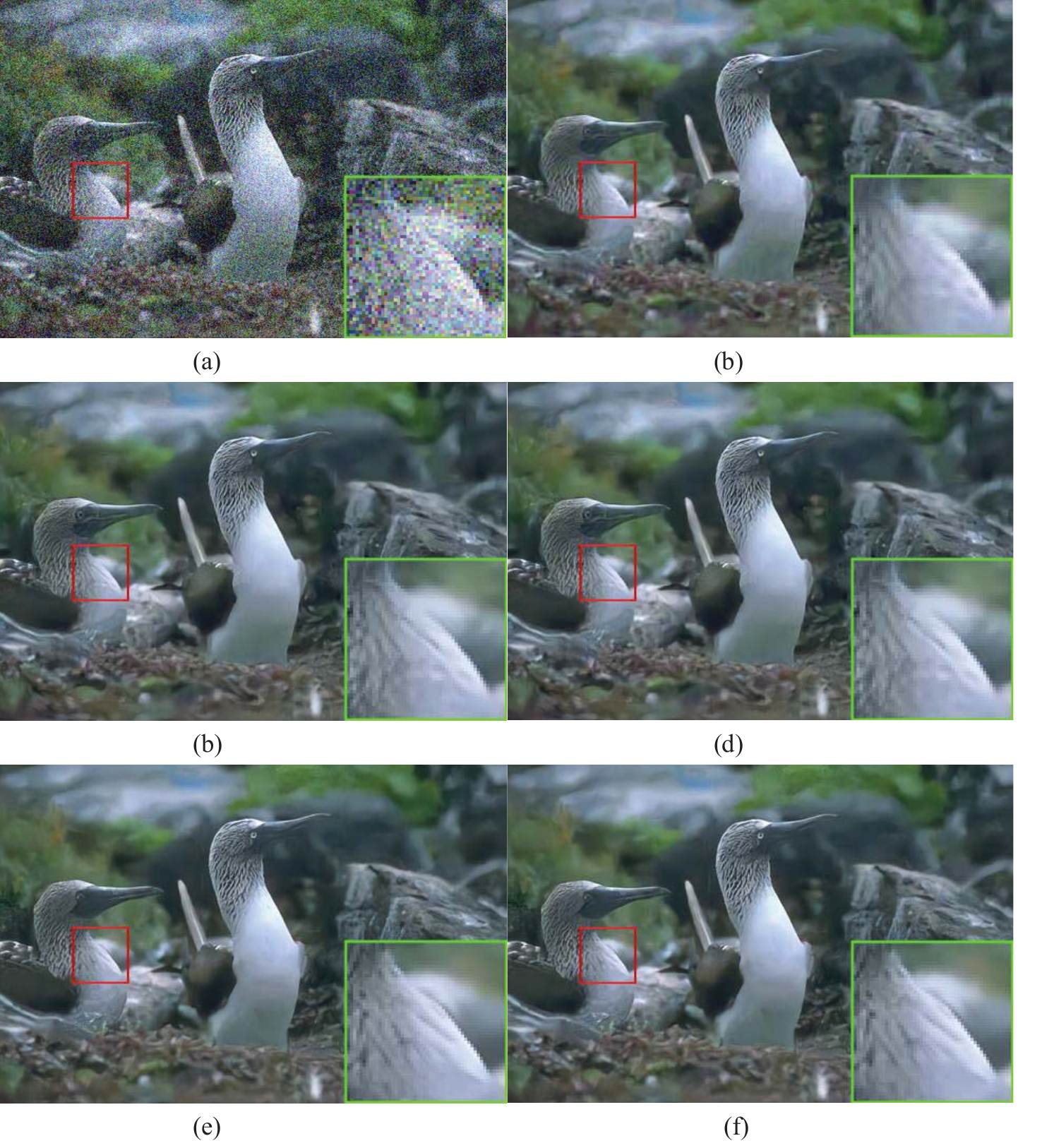}
\label{Fig:a}}
\hfil
\caption{Visual comparison for a color Gaussian noisy image from CBSD68 obtained by different methods, when noise level is 35. (a) Noisy image/17.48dB, (b) BM3D/30.52dB, (c) IRCNN/31.00dB, (d) FFDNet/31.00dB, DudeNet/31.13dB and (e) DudeNet-B/31.08dB.}
\end{figure}
\begin{figure}[!htbp]
\centering
\subfloat{\includegraphics[width=3.6in]{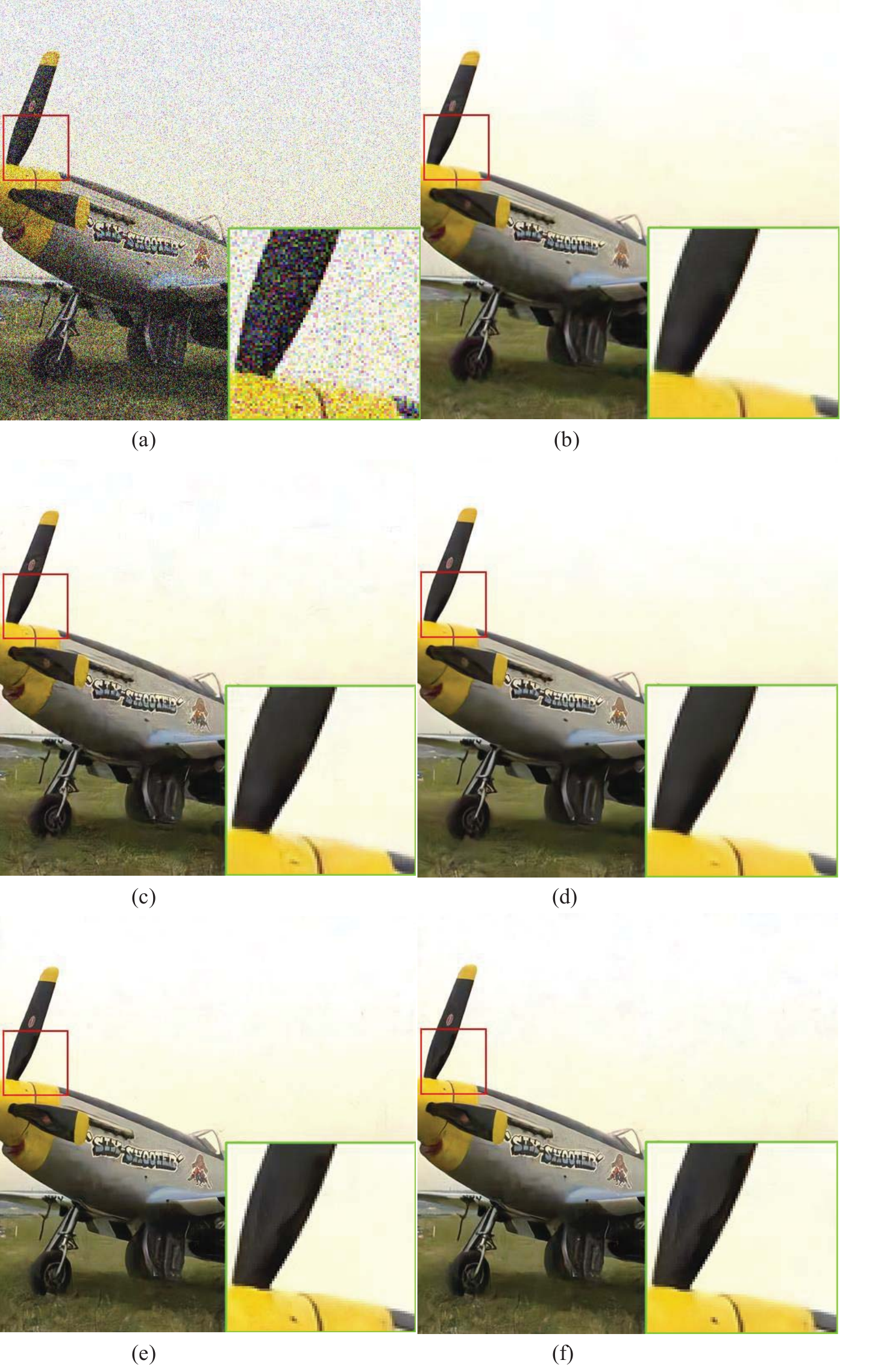}
\label{Fig:a}}
\hfil
\caption{Visual results for a color Gaussian noisy image  from Kodak24 obtained by different methods, when noise level is 50. (a) Noisy image/14.28dB, (b) BM3D/30.33dB, (c) IRCNN/30.93B, (d) FFDNet/31.15dB, (e) DudeNet/31.37dB and (d) DudeNet-B/31.26dB.}
\end{figure}
\subsection{Comparisons with state-of-the-art denoising methods}
In this paper, we conduct comparative experiments for four applications: gray and color synthetic noisy images, blind denoising, and real noisy images, and compare the run-time of each image. For these experiments, we choose state-of-the-art methods, including BM3D \cite{dabov2007image}, WNNM \cite{gu2014weighted}, expected patch log likelihood (EPLL) \cite{zoran2011learning}, multi-layer perceptron (MLP) \cite{burger2012image}, CSF \cite{schmidt2014shrinkage}, TNRD \cite{chen2016trainable}, DnCNN \cite{zhang2017beyond}, DnCNN for blind denoising (DnCNN-B)\cite{zhang2017beyond}, IRCNN \cite{zhang2017learning}, FFDNet \cite{zhang2018ffdnet}, targeted image denoising (TID) \cite{luo2015adaptive}, GAT-BM3D \cite{makitalo2012optimal}, MemNet \cite{tai2017memnet}, RED30 \cite{mao2016image}, and enhanced convolutional neural denoising network (ECNDNet) \cite{tian2019enhanced}. Specifically, we use PSNR \cite{zhang2018learning,huang2013self} and run-time are  performance metrics for comparing the denoising methods: $PSNR{\rm{ = }}10 \times {\rm{lo}}{{\rm{g}}_{10}}(\frac{{{{(MA{X})}^2}}}{{MSE}})$, where $MAX$ and $MSE$ are the maximum pixel value and mean squared error between a given clean and predicted clean images, respectively.
\begin{table}[t!]
\caption{Average PSNR (dB) values for different methods on CBSD68 and Kodak24, with noise levels of 15, 25, 35, 50, 75 and varying noise levels [0,55].}
\label{tab:1}
\centering
\scalebox{0.87}[0.87]{
\begin{tabular}{|c|c|c|c|c|c|c|}
\hline
Datasets & Methods & $\sigma$ = 15 & $\sigma$ = 25 & $\sigma$ = 35 & $\sigma$ = 50 & $\sigma$ = 75 \\
\hline
\multirow{6}{*}{CBSD68}
&CBM3D \cite{dabov2007image} &33.52 &30.71 &28.89 &27.38 &25.74\\
\cline{2-7} &FFDNet \cite{zhang2018ffdnet} &33.80 &31.18 &29.57 &27.96  &\textcolor{blue}{26.24} \\
\cline{2-7} &DnCNN \cite{zhang2017beyond} &\textcolor{blue}{33.98} &31.31   &29.65 &28.01 &- \\
\cline{2-7} &IRCNN \cite{zhang2017learning} &33.86 &31.16 &29.50 &27.86 &-\\
\cline{2-7} &DudeNet &\textcolor{red}{34.01}  &\textcolor{red}{31.34}   &\textcolor{red}{29.71} &\textcolor{red}{28.09} &\textcolor{red}{26.40}\\
\cline{2-7} &DudeNet-B &33.96  &\textcolor{blue}{31.32}   &\textcolor{blue}{29.69} &\textcolor{blue}{28.05} &-\\
\hline
\multirow{6}{*}{Kodak24}
&CBM3D \cite{dabov2007image} &34.28 &31.68 &29.90 &28.46 &26.82\\
\cline{2-7} &FFDNet \cite{zhang2018ffdnet} &34.55 &32.11 &30.56 &28.99  &\textcolor{blue}{27.25} \\
\cline{2-7} &DnCNN  \cite{zhang2017beyond} &\textcolor{blue}{34.73} &\textcolor{blue}{32.23}   &30.64 &29.02 &- \\
\cline{2-7} &IRCNN \cite{zhang2017learning} &34.56 &32.03 &30.43 &28.81 &-\\
\cline{2-7}  &DudeNet &\textcolor{red}{34.81}  &\textcolor{red}{32.26}   &\textcolor{red}{30.69} &\textcolor{red}{29.10} &\textcolor{red}{27.39}\\
\cline{2-7}  &DudeNet-B &34.71  &\textcolor{blue}{32.23}   &\textcolor{blue}{30.66} &\textcolor{blue}{29.05} &-\\
\hline
\end{tabular}}
\end{table}
\subsubsection{Gray and color synthetic noisy images}
Table IV illustrates the average PSNR on the benchmark dataset BSD68, for gray synthetic noisy images. As can be seen, DudeNet outperforms several state-of-the-art denoisers, i.e., DnCNN and FFDNet. Furthermore, DudeNet with blind denoising (DudeNet-B) also obtains a good performance. For example, DudeNet-B has improvements of 0.02dB over DNCNN when $\sigma  = 50$. Fig. 3 illustrates the visual images from BM3D, IRCNN, FFDNet, DudeNet and DudeNet-B on BSD68. Table V shows the good denoising performance of DudeNet for each category of gray synthetic noisy images, where it obtains the best denoising result for various noise levels (i.e., 15, 25, 50). Fig. 4 shows final images. As can be seen, DudeNet produces a much clearer image than IRCNN. From Table VI, at different noise levels (i.e., 15, 25, 35, 50, 75, [0, 55]), DudeNet and DudeNet-B is very competitive with other popular methods for color synthetic noisy images from CBSD68 and Kodak24. Figs. 5 and 6 illustrates visual images.
\begin{table}[t!]
\caption{Average PSNR (dB) values for different methods from public dataset cc.}
\label{tab:1}
\centering
\scalebox{0.65}[0.70]{
\begin{tabular}{|c|c|c|c|c|c|c|}
\hline
Camera Settings &GAT-BM3D \cite{makitalo2012optimal} &CBM3D \cite{dabov2007image}  &TID \cite{luo2015adaptive} &CSF \cite{schmidt2014shrinkage} &DnCNN \cite{zhang2017beyond} &DudeNet \\
\hline
\multirow{3}{*}{Canon 5D ISO=3200}
&31.23 &\textcolor{red}{39.76} &37.22 &35.68 &\textcolor{blue}{37.26}  &36.66 \\
\cline{2-7} &30.55 &\textcolor{blue}{36.40} &34.54 &34.03 &34.87  &\textcolor{red}{36.70} \\
\cline{2-7} &27.74 &\textcolor{red}{36.37} &34.25 &32.63 &34.09   &\textcolor{blue}{35.03} \\
\hline
\multirow{3}{*}{Nikon D600 ISO=3200}
&28.55 &\textcolor{red}{34.18} &32.99 &31.78 &33.62    &\textcolor{blue}{33.72} \\
\cline{2-7} &32.01 &\textcolor{blue}{35.07} &34.20  &\textcolor{red}{35.16} &34.48  &34.70 \\
\cline{2-7} &\textcolor{blue}{39.78} &37.13 &35.58 &\textcolor{red}{39.98} &35.41  &37.98\\
\hline
\multirow{3}{*}{Nikon D800 ISO=1600}
&32.24 &36.81 &34.49 &34.84 &\textcolor{blue}{37.95}   &\textcolor{red}{38.10}\\
\cline{2-7} &33.86 &37.76 &35.19 &\textcolor{blue}{38.42} &36.08 &\textcolor{red}{39.15}\\
\cline{2-7} &33.90 &\textcolor{red}{37.51} &35.26 &35.79 &35.48  &\textcolor{blue}{36.14}\\
\hline
\multirow{3}{*}{Nikon D800 ISO=3200}
&36.49 &35.05 &33.70 &\textcolor{red}{38.36} &34.08   &\textcolor{blue}{36.93}\\
\cline{2-7} &32.91 &34.07 &31.04 &\textcolor{blue}{35.53} &33.70   &\textcolor{red}{35.80}\\
\cline{2-7} &\textcolor{red}{40.20} &34.42 &33.07 &\textcolor{blue}{40.05} &33.31   &37.49\\
\hline
\multirow{3}{*}{Nikon D800 ISO=6400}
&29.84 &31.13 &29.40 &\textcolor{red}{34.08} &29.83   &\textcolor{blue}{31.94}\\
\cline{2-7} &27.94 &31.22 &29.86 &\textcolor{blue}{32.13} &30.55   &\textcolor{red}{32.51}\\
\cline{2-7} &29.15 &30.97 &29.21 &\textcolor{blue}{31.52} &30.09  &\textcolor{red}{32.91}\\
\hline
Average &32.43 &35.19 &33.36 &\textcolor{blue}{35.33} &33.86  &\textcolor{red}{35.72}\\
\hline
\end{tabular}}
\end{table}
\begin{table}[t!]
\label{tab:1}
\caption{Run-time comparison of nine denoising methods for the noisy images of sizes 256 $\times$ 256, 512 $\times$ 512 and 1024 $\times$ 1024.}
\centering
\scalebox{0.8}[0.8]{
\begin{tabular}{|c|c|c|c|c|}
\hline
Methods &Device &256 $\times$ 256 &512 $\times$ 512 &1024 $\times$ 1024\\
\hline
BM3D \cite{dabov2007image} &CPU &0.59  &2.52 &10.77 \\
\hline
WNNM \cite{gu2014weighted} &CPU &203.1  &773.2 &2536.4 \\
\hline
EPLL \cite{zoran2011learning} &CPU &25.4  &45.5 &422.1 \\
\hline
MLP \cite{burger2012image} &CPU &1.42  &5.51 &19.4 \\
\hline
TNRD \cite{chen2016trainable} &CPU &\textcolor{blue}{0.45}  &1.33 &4.61 \\
\hline
CSF \cite{schmidt2014shrinkage} &GPU &- &\textcolor{blue}{0.92} &\textcolor{blue}{1.72} \\
\hline
RED30 \cite{mao2016image} &GPU &1.362 &4.702 &15.77\\
\hline
MemNet \cite{tai2017memnet} &GPU &0.878 &3.606 &14.69\\
\hline
DudeNet &GPU &\textcolor{red}{0.018}  &\textcolor{red}{0.422} &\textcolor{red}{1.246}\\
\hline
\end{tabular}}
\label{tab:booktabs}
\end{table}
\subsubsection{Blind denoising}
The blind denoising model DudeNet-B is trained from 0 to 55. As shown in Table IV-VI, we can see that DudeNet-B is very competitive to the FFDNet and IRCNN for gray and color images denoising. That proves that our model is robust for blind denoising.
\subsubsection{Real noisy images}
Table VII shows denoising performances for real noisy images. DudeNet achieves excellent results, with an improvement of 1.86dB over DnCNN. In a summary, our denoising model is suitable to complex noisy tasks, such as color synthetic noisy images, blind denoising and real noisy images.

Table VIII shows the running times of the nine methods on each different sized image, where DudeNet is more competitive than the state-of-the-art denoisers, i.e., RED30 and MemMet. Tables IV-VIII illustrate the denoising performance (i.e., PSNR and running time) of the different methods, where red and blue lines represent the best and second best results for image denoising, respectively.

According to the previous analysis in Section IV. D and experiment verification in Section IV. E, we can refine the merits of this paper as follows.

Firstly, dual networks with sparse mechanism can extract different features to boost the generalized ability of denosier for addressing complex tasks, e.g. real noisy image and blind noise.

Secondly, combining global and local features can obtain salient features to recover fine details, which can consolidate dual networks to resolve complex tasks.

Finally, small filter size is used to reduce the complexity of the denoiser.

\section{Conclusion}
In this paper, we proposed a novel DudeNet for image denoising. DudeNet uses dual networks to
extract diverse features for boosting the representing ability of the learned features for denoising. The sparse mechanism of DudeNet can help well trade denoising performance for processing speed by extracting both global and local features to fuse them to obtain salient features to recover fine details for complex noisy images. We also proposed using compression blocks  to reduce redundant information so as to reduce computational cost and memory consumption. Extensive experiments demonstrate of high visual quality and computation efficiency of DudeNet. In the future, we intend to extend DudeNet to deal with multiple low-level vision tasks, including image super-resolution and deblurring.




\ifCLASSOPTIONcaptionsoff
  \newpage
\fi



%
\bibliographystyle{IEEEtran}
\bibliography{tcw}
\begin{IEEEbiography}[{\includegraphics[width=0.9in,height=1.2in]{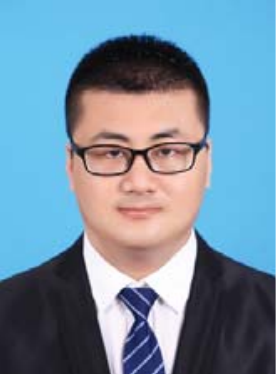}}]{Chunwei Tian}
is currently pursuing the Ph.D degree in the School of Computer Science and Technology at Harbin Institute of Technology, Shenzhen. He received the B.S. degree and M.S. degree at Harbin University of Science and Technology, in 2014 and 2017, respectively. His research interests include image denoising, image super-resolution, pattern recognition and deep learning. He has published over 20 papers in academic journals and conferences, including IEEE Transactions on Multimedia, Neural Networks, Pattern Recognition Letters, ICASSP and ICPR. He is a PC of the 18th IEEE International Conference on Dependable, Autonomic and Secure Computing (DASC 2020), a PC Assistant of IJCAI 2019, a reviewer of some journals and conferences, such as the IEEE Transactions on Image Processing, the IEEE Transactions on Industrial Informatics, the IEEE Transactions on Systems, Man and Cybernetics: Systems, the Computer Vision and Image Understanding, the Nerocomputing, the Visual Computer, the Journal of Modern Optics, the IEEE Access, the CAAI Transactions on Intelligence Technology, the International Journal of Biometrics, the International Journal of Image and Graphics, the ACAIT2019 and the AIIPC 2019.
\end{IEEEbiography}
\begin{IEEEbiography}[{\includegraphics[width=1.0in,height=1.2in]{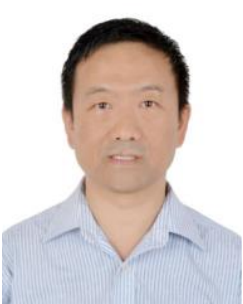}}]{Yong Xu}
(Senior Member, IEEE)
 received his B.S. degree, M.S. degree in 1994 and 1997, respectively. He received the Ph.D. degree in Pattern Recognition and Intelligence system at NUST (China) in 2005. Now he works at Harbin Institute of Technology, Shenzhen. His current interests include pattern recognition, deep learning, biometrics, machine learning and video analysis. He has published over 100 papers in toptier academic journals and conferences. He has served as an Co-Editors-in-Chief of the International Journal of Image and Graphics, an Associate Editor of the CAAI Transactions on Intelligence Technology, an editor of the Pattern Recognition and Artificial Intelligence. More information please refer to http://www.yongxu.org/lunwen.html.

\end{IEEEbiography}
\begin{IEEEbiography}[{\includegraphics[width=0.9in,height=1.2in]{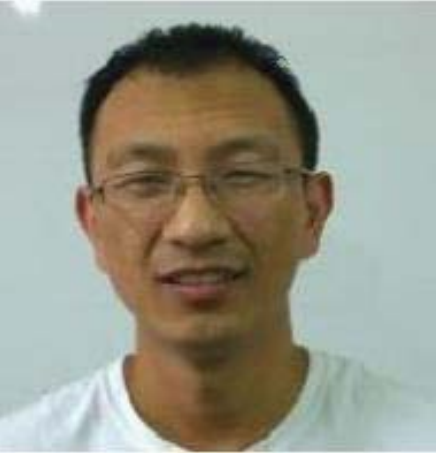}}]{Wangmeng Zuo}
(Senior Member, IEEE) received the Ph.D. degree in computer application technology from the Harbin Institute of Technology, Harbin, China, in 2007. He is currently a Professor in the School of Computer
Science and Technology, Harbin Institute of Technology. His current research interests include image enhancement and restoration, object detection, visual tracking, and image classification.
He has published over 100 papers in toptier academic journals and conferences. He has served as a Tutorial Organizer in ECCV 2016, an Associate Editor of the IET Biometrics and Journal of Electronic Imaging, and the Guest Editor of Neurocomputing, Pattern Recognition, IEEE Transactions on Circuits and Systems for Video Technology, and IEEE Transactions on Neural Networks and Learning Systems.
\end{IEEEbiography}
\begin{IEEEbiography}[{\includegraphics[width=0.9in,height=1.2in]{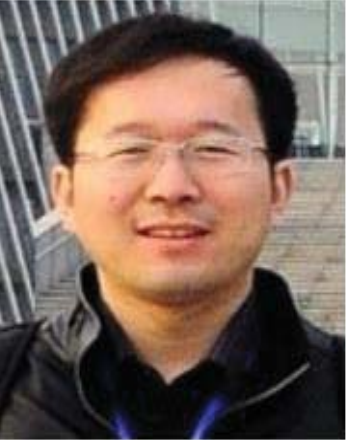}}]{Bo Du}
(Senior Member, IEEE) received the B.S. and Ph.D. degrees in photogrammetry and remote sensing from the State Key Laboratory of Information
Engineering in Surveying, Mapping, and Remote Sensing, Wuhan University, Wuhan, China, in 2005 and 2010, respectively.

He is currently a Professor with the School of Computer, Wuhan University. He was with the Centre for Quantum Computation and Intelligent
Systems, University of Technology Sydney, Ultimo, NSW, Australia. He has authored over 40 research papers in the IEEE Transactions on Geoscience and Remote Sensing, the IEEE Transactions on Image
Processing, the IEEE Journal of Selected Topics in Applied Earth Observations and Remote Sensing, and the IEEE Geoscience and Remote Sensing Letters. His current research interests include
pattern recognition, hyperspectral image processing, and signal processing.

Dr. Du was a recipient of the Best Reviewer Award from the IEEE
Geoscience and Remote Sensing Society for his services to the IEEE
JSTARS in 2011 and the ACM Rising Star Award for his academic
progress in 2015. He was the Session Chair for the IEEE IGARSS
2016 and 4th IEEE GRSS Workshop on Hyperspectral Image and Signal
Processing: Evolution in Remote Sensing.  He has served as Associate Editor of the Neurocomputing and IEEE Access.
\end{IEEEbiography}

\begin{IEEEbiography}[{\includegraphics[width=0.9in,height=1.2in]{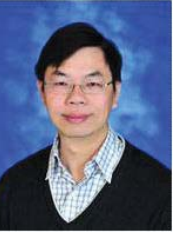}}]{Chia-Wen Lin}
(Fellow, IEEE) received the Ph.D degree in electrical engineering from National Tsing Hua University (NTHU), Hsinchu, Taiwan, in 2000.
He is currently Professor with the Department of Electrical Engineering and the Institute of Communications Engineering, NTHU. He is also Deputy Director of the AI Research Center of NTHU.  His research interests include computer vision and image and video processing.
He has served as a Distinguished Lecturer for IEEE Circuits and Systems Society from 2018 to 2019, a Steering Committee member for the \textsc{IEEE Transactions on Multimedia} from 2014 to 2015, and the Chair of the Multimedia Systems and Applications Technical Committee of the IEEE Circuits and Systems Society from 2013 to 2015.	His articles received the Best Paper Award of IEEE VCIP 2015, Top 10\% Paper Awards of IEEE MMSP 2013, and the Young Investigator Award of VCIP 2005.   He has been serving as President of the Chinese Image Processing and Pattern Recognition Association, Taiwan.  He served as a General Co-Chair of IEEE VCIP 2018, a Technical Program Co-Chair of IEEE ICME 2010,  and a Technical Program Co-Chair of IEEE ICIP 2019. He  has served as an Associate Editor for the \textsc{IEEE Transactions on Image Processing}, the \textsc{IEEE Transactions on Circuits and Systems for Video Technology}, the \textsc{IEEE Transactions on Multimedia}, \textsc{IEEE Multimedia}, and the \textit{Journal of Visual Communication and Image Representation}.
\end{IEEEbiography}

\begin{IEEEbiography}[{\includegraphics[width=0.9in,height=1.2in]{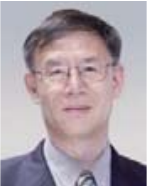}}]{David Zhang}
(Life Fellow, IEEE) received the Graduation degree
in computer science from Peking University, Beijing, China, the M.Sc. degree in computer science
and the Ph.D. degree from the Harbin Institute of
Technology (HIT), Harbin, China, in 1982 and 1985,
respectively, and the second Ph.D. degree in electrical and computer engineering from the University of
Waterloo, Waterloo, ON, Canada, in 1994.
He is currently a Chair Professor with The
Hong Kong Polytechnic University, Hong Kong, and
The Chinese University of Hong Kong (Shenzhen),
Shenzhen, China. He also serves as a Visiting Chair Professor with Tsinghua
University, Beijing, China, and an Adjunct Professor with Peking University,
Beijing, China, Shanghai Jiao Tong University, Shanghai, China, HIT, and the University of Waterloo. His research interests include image processing, pattern recognition and biometrics.

Dr. Zhang is a Life fellow of IEEE and IAPR fellow. And he is also a Croucher Senior Research Fellow, Distinguished Speaker of the IEEE Computer Society.
He is the Founder and Editor-in-Chief, International Journal of Image and Graphics (IJIG); Book Editor, Springer International Series on Biometrics (KISB); Organizer, the first International Conference on Biometrics Authentication (ICBA); Associate Editor of more than ten international journals including IEEE Transactions on Systems, Man, and Cybernetics: Systems and so on. Over past 30 years, he have been working on pattern recognition, image processing and biometrics, where many research results have been awarded and some created directions, including palmprint recognition, computerized TCM and facial beauty analysis, are famous in the world. So far, he has published over 20 monographs, 400 international journal papers and 40 patents from USA/Japan/HK/China. He has been continuously listed as a Highly Cited Researchers in Engineering by Clarivate Analytics (formerly known as Thomson Reuters) in 2014, 2015, 2016, 2017, 2018 and 2019, respectively (http://highlycited.com). He is also ranked 80 with H-Index 108 at Top 1000 Scientists for international Computer Science and Electronics (http://www.guide2research.com/scientists).
\end{IEEEbiography}
\end{document}